\documentclass[conference]{IEEEtran}

\usepackage{cite}

\usepackage{cite}
\usepackage{amsmath,amssymb,amsfonts}
\usepackage{algorithmic}
\usepackage{graphicx}
\usepackage{textcomp}
\usepackage{xcolor}
\def\BibTeX{{\rm B\kern-.05em{\sc i\kern-.025em b}\kern-.08em
    T\kern-.1667em\lower.7ex\hbox{E}\kern-.125emX}}

\usepackage{placeins}
\usepackage{float}
\usepackage{subcaption}

\usepackage{authblk}
\usepackage{caption}
\usepackage{siunitx}
\makeatletter

\newcounter{author}
\renewcommand{\author}[2][]{
   \stepcounter{author}
   \@namedef{author@\theauthor}{#2}
   \@namedef{authorlabel@\theauthor}{#1}
}

\newcounter{address}
\newcommand{\address}[2][]{
   \stepcounter{address}
   \@namedef{address@\theaddress}{#2}
   \@namedef{addresslabel@\theaddress}{#1}
}

\newcommand{\alsep}{and}

\def\newmaketitle{\par%
  \begingroup%
  \normalfont%
  \def\thefootnote{}
  \def\footnotemark{}
  \let\@makefnmark\relax
  \footnotesize
  \footnotesep 0.7\baselineskip
  \normalsize%
  \twocolumn[\thenewmaketitle\@IEEEaftertitletext]%
  \if@IEEEusingpubid
     \enlargethispage{-\@IEEEpubidpullup}%
  \fi
  \endgroup
  \setcounter{footnote}{0}\let\maketitle\relax\let\@maketitle\relax
  \gdef\@thanks{}%
  \let\thanks\relax}

\def\thenewmaketitle{
  \newpage
  \begin{center}%
    \vskip0.2em{\Huge\@IEEEcompsoconly{\sffamily}\@IEEEcompsocconfonly{\normalfont\normalsize\vskip 2\@IEEEnormalsizeunitybaselineskip
   \bfseries\large}\@title\par}\vskip1.0em\par%
    \vspace{1ex}
    \newcounter{c@author}
    \newcounter{c@tmp}
    \ifthenelse{\value{author}=2}{%
      \newcommand{\liand}{ and }}{%
      \newcommand{\liand}{, and }}
    \ifthenelse{\value{address}<2}{%
      \@nameuse{author@1}%
      \stepcounter{c@author}%
      \whiledo{\value{c@author}<\value{author}}{%
        \setcounter{c@tmp}{\value{author}}%
        \addtocounter{c@tmp}{-\value{c@author}}%
        \ifthenelse{\value{c@tmp}=1}{%
          \renewcommand{\alsep}{\liand}}{\renewcommand{\alsep}{, }}%
        \stepcounter{c@author}\alsep \@nameuse{author@\thec@author}}\\%
    }
    {
      \@nameuse{author@1}${}^{(\ref{\@nameuse{authorlabel@1}})}$%
      \stepcounter{c@author}%
      \whiledo{\value{c@author}<\value{author}}{%
      \setcounter{c@tmp}{\value{author}}%
      \addtocounter{c@tmp}{-\value{c@author}}%
      \ifthenelse{\value{c@tmp}=1}{%
        \renewcommand{\alsep}{\liand}}{\renewcommand{\alsep}{, }}%
      \stepcounter{c@author}\alsep \@nameuse{author@\thec@author}%
        ${}^{(\ref{\@nameuse{authorlabel@\thec@author}})}$%
      }
    }
    \vspace{0.2ex}

    \ifthenelse{\value{address}>0}{%
      \ifthenelse{\value{address}=1}{
        {\@nameuse{address@1}}
      }
      {
        \newcounter{c@address}

        \begin{center}
        \whiledo{\value{c@address}<\value{address}}
        {
          \refstepcounter{c@address}
            ${}^{(\thec@address)}$\,%
              \label{\@nameuse{addresslabel@\thec@address}}%
              \@nameuse{address@\thec@address}\\ %
        }
        \end{center}
      } 
    }
    {
      \relax
    }
  \end{center}
}

\makeatother

\title{Electrical and Thermal Performance Tuning of Spoof Plasmonic Interconnect}

\author[org1]{Rafichha Yasmin}
\author[org2]{Ishrat Jahan}
\author[org2]{Abdelrahman Omar}
\author[org1]{Md. Zunaid Baten}
\author[org1]{A.B.M. Harun-ur Rashid}
\author[org2]{Soumitra Joy\textsuperscript{*}}

\address[org1]{Bangladesh University of Engineering and Technology, Dhaka, Bangladesh}
\address[org2]{University of North Carolina at Charlotte, Charlotte, NC 28223, USA (\textsuperscript{*}{s.joy@charlotte.edu})}

\begin{document}

\newmaketitle

\begin{abstract}
This work introduces an electromagnetic metastructure based interconnect design that could address the critical need for electrical bandwidth and heat dissipation in high-speed, chiplet integration. We leverage silicon as the substrate for its superior thermal properties, and to counteract its high dielectric constant that typically causes high mutual capacitance among interconnects, we've engineered a periodically corrugated, compact metallic structure enabling signal propagation via strongly confined spoof surface plasmon polaritons (SSPPs). By placing this engineered metal on a $50$ $\mu$m oxide layer atop Si substrate, we achieved a low insertion loss of $0.015$ dB/cm and a $10$ dB reduction in crosstalk noise within $5$ GHz, resulting in a bandwidth $2.5\times$ as high as that of a standard microstriplines of the same footprint. Furthermore, a $5$ ns input pulse showed minimal distortion and a $0.13$ ns/cm propagation delay in our proposed interconnect. Critically, the thin oxide layer minimally impacted the heat dissipation of Si substrate, demonstrating a fourfold reduction in temperature compared to an FR4 substrate. These full-wave simulation-supported findings present a viable pathway for high-density, thermally efficient interconnects in advanced packaging.
\end{abstract}

\section{Introduction}

As heterogeneous integration scales computing speed, high-performance interconnect design becomes critical. It links multiple chip-lets on a single package, with bandwidth, signal integrity, and thermal integrity dictating overall system performance\cite{b1}.The substrate plays a crucial role in interconnect design. It needs high electrical insulation to prevent leakage, and also substantial thermal conductivity for heat dissipation. This dual demand is inherently contradictory for most materials, as the same charge carriers facilitate both electrical and thermal transport\cite{b2}. For instance, while low-k materials excel electrically in high-speed interconnects (boosting bandwidth, reducing leakage), their thermal conduction is orders of magnitude worse than silicon\cite{b3}. This fundamental dichotomy necessitates novel interconnect designs that offer enhanced control to independently optimize both electrical and thermal characteristics.

This work proposes electromagnetic meta-structures on oxide coated silicon substrate for interconnect design, providing more flexibility in controlling electrical and thermal performance. We specifically leverage spoof surface plasmon polariton (SSPP) mode \cite{b4} for better electrical control and silicon substrate for better thermal property. Our previous research demonstrated that SSPP waveguides enhance bandwidth in high-density interconnects by minimizing crosstalk\cite{b5}. This work advances the field of spoof plasmonic interconnect by:
\begin{itemize}
\item Proposing a compact, periodic meandered metal structure on silicon that supports highly confined SSPP modes across a wide frequency band. This design achieves twice the transmission bandwidth and a 10 dB lower crosstalk noise floor compared to standard microstrip-lines.
\item Demonstrating that integrating a thin oxide layer between the metal and silicon substrate in these meandered SSPP waveguides significantly minimizes electromagnetic signal insertion loss, while retaining the benefit of better heat dissipation compared to low-k substrate.

\end{itemize}

\section{Theory and Device Structure}

\begin{figure}[ht]
    \centering
    \includegraphics[width=0.7\linewidth]{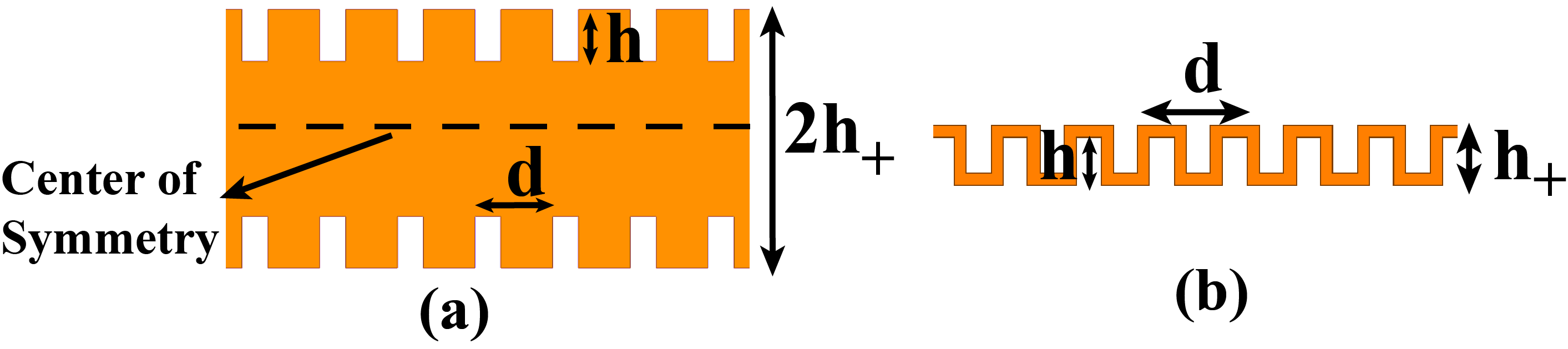}
    \caption{(a) A double sided, mirror symmetric SSPP waveguide  (b) A compact design of double sided SSPP waveguide proposed in this work, we name it as meandered SSPP}
    \label{fig:sspp1}
\end{figure}

A spoof surface plasmon polariton (SSPP) mode arises from the interaction between free-space waves and quarter-wavelength resonant cavities formed by periodic corrugations on a metal strip. The closely spaced resonant cavities couple to each other, creating a propagating mode that adheres to the metal surface. In this work, we propose `meandered SSPP waveguide', a variant of the typical double-sided SSPP interconnect shown in Fig.~\ref{fig:sspp1}. In their standard design, a metal strip has both of its edge periodically corrugated with indentation depth $h$ and period $d$, making it mirror-symmetric about the long axis of the strip (Fig.~\ref{fig:sspp1}(a)). In the meandered design, corrugations on one edge is translated by half a period ($d/2$) relative to the other edge. The corrugation width ($a$) is less than $d/2$, allowing the two comb-shaped edges to interlock(Fig.~\ref{fig:sspp1}(b)). This configuration offers two main benefits: 1) compact size: the minimum waveguide width is reduced to $h_+$, 2) stronger coupling: adjacent corrugations couple more strongly, leading to better mode confinement. To ensure the waveguide supports DC signals, a ground plane is added beneath the substrate, as SSPP mode confinement weakens significantly at DC frequencies. Two substrate materials were evaluated, both with a thickness of $2.6$ mm in order to match the typical thickness of commercially available FR4 board for PCB design: 1) Low-k FR4 epoxy with dielectric constant $\epsilon_r=4.4$, and 2) High-k silicon with $\epsilon_r=11.9$.

\begin{figure}[ht]
    \centering
    \includegraphics[width=0.85\linewidth]{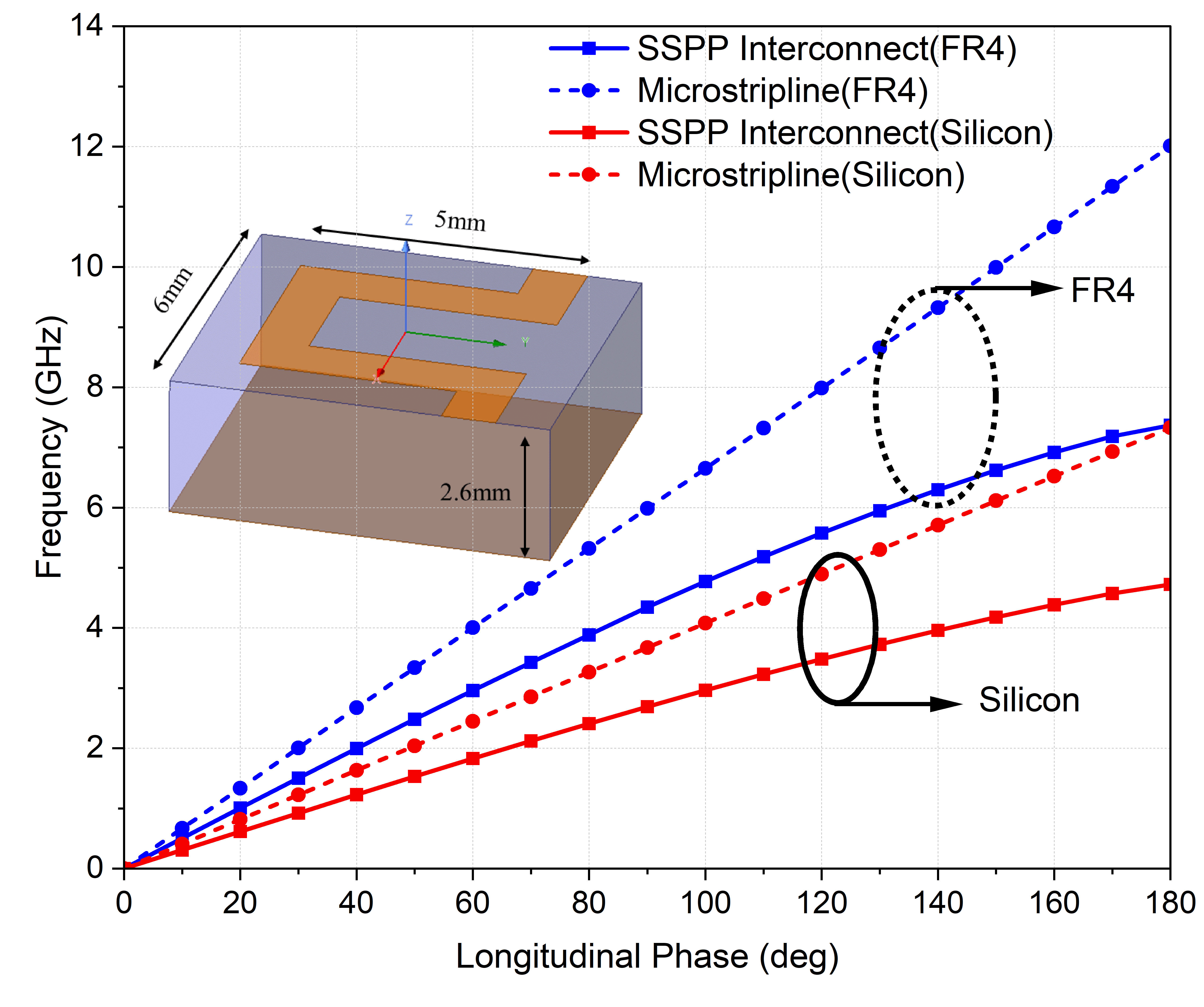}
    \caption{Frequency vs longitudinal phase dispersion relation for a unit cell of SSPP metal wire with ground plane, shown in the inset, and that for a microstripline for two different substrate materials: FR4 Epoxy and Silicon.}
    \label{fig:dispersion}
\end{figure}

\begin{figure}[ht]
    \centering
    \includegraphics[width=0.8\linewidth]{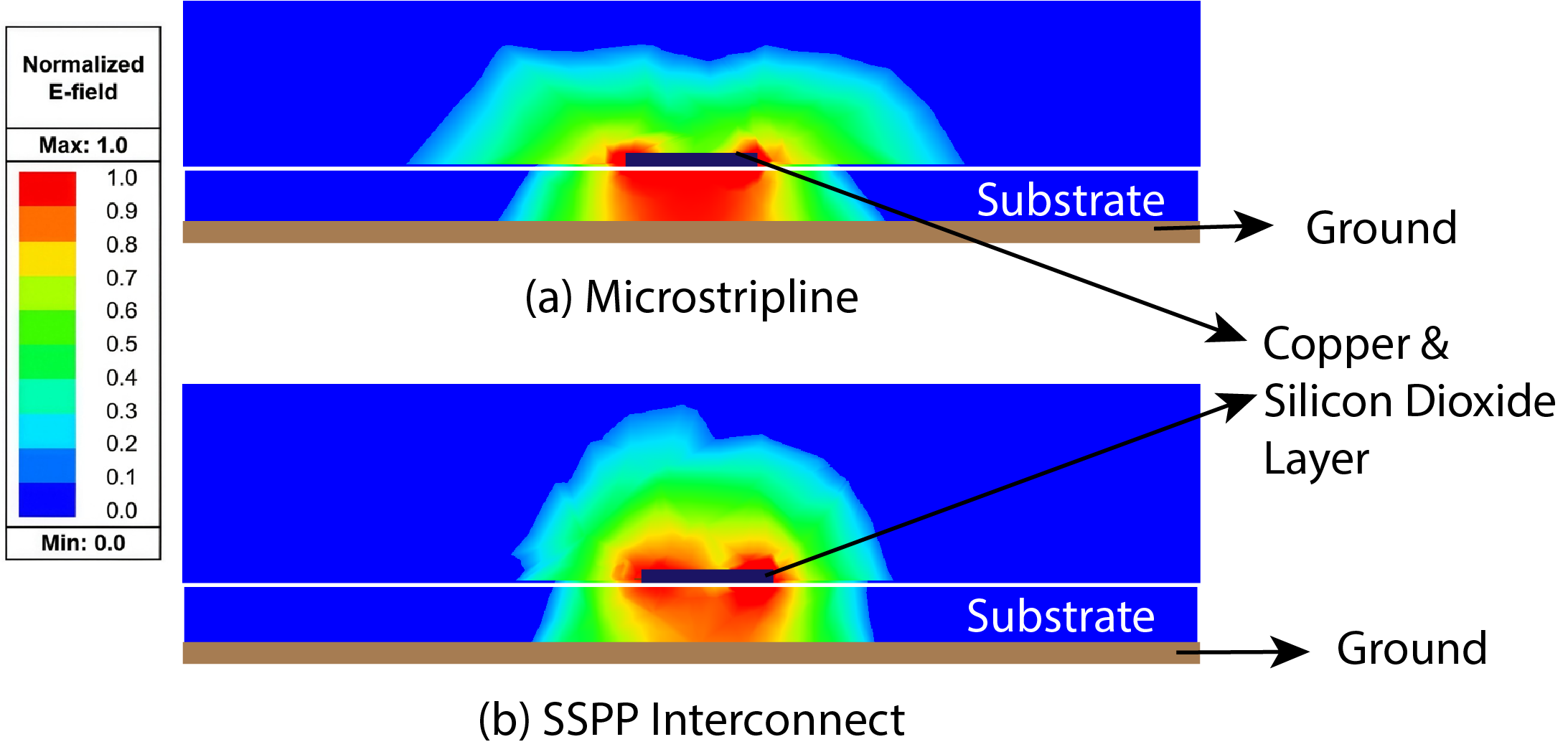}
    \caption{Cross-Sectional Electric Field Profiles of a (a) standalone microstripline, and (b) standalone meandered SSPP Interconnects at $4$GHz}
    \label{fig:E-field_cross-section}
\end{figure}

\begin{figure}[ht]
    \centering
    \includegraphics[width=0.85\linewidth]{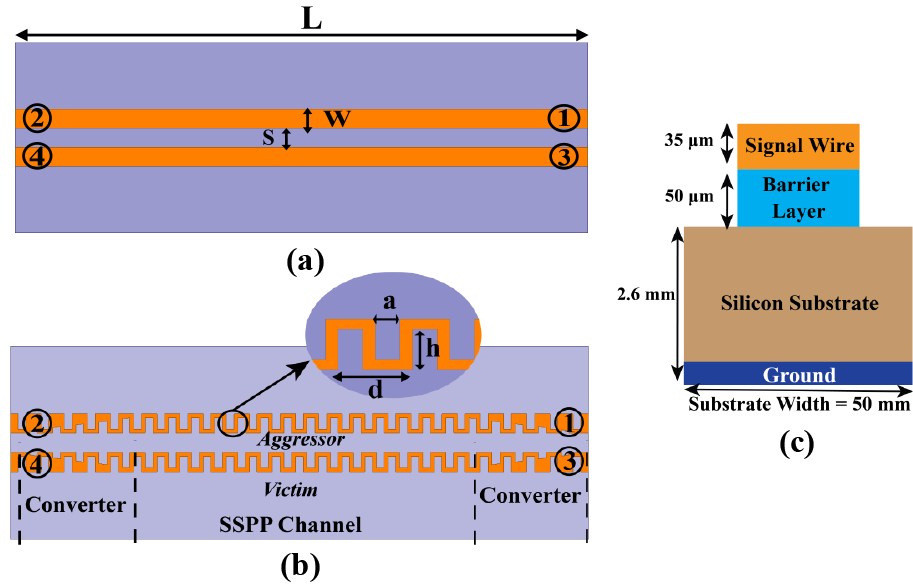}
    \caption{(a) Geometric features of two adjacent Microstriplines (b) Geometric features of two adjacent meandered SSPP Interconnects (c) Cross-sectional geometry of interconnect with intermediate oxide layer on silicon substrate}
    \label{fig:Geometry}
\end{figure}

Fig.~\ref{fig:dispersion} illustrates the dispersion relation (frequency $f$ vs. wavenumber $k=\theta/d$) for the proposed meandered SSPP waveguide, comparing it to a microstripline of equal width ($W=5$ mm). Both FR4 and silicon substrates show significantly stronger mode confinement in the meandered SSPP waveguide. This enhanced confinement suggests greater immunity to crosstalk noise in multi-interconnect scenarios. Notably, the cut-off frequency of the SSPP waveguide decreases from 7 GHz (FR4) to 4.8 GHz (silicon). This is because the cut-off frequency is inversely proportional to the optical length of the corrugation, $h_{op}=\sqrt{\epsilon}h$, Consequently, low-k materials offer higher SSPP waveguide bandwidth, though they may present a thermal trade-off, which will be discussed later.
Fig.~\ref{fig:E-field_cross-section}a,b display the electric field distribution across a cross-section of a microstripline and an meandered SSPP waveguide, respectively, on silicon substrate at $4$ GHz. A visual comparison clearly shows that the electric field in the SSPP waveguide remains more tightly confined to the top metal surface than in the microstripline.
Fig.~\ref{fig:Geometry}(a) displays the configuration of the interconnect that has been considered as a reference device, while Fig.~\ref{fig:Geometry}(b) represents our proposed device with improved performance. The reference device consists of two plain metal microstriplines, whereas the proposed device is engineered, meandered SSPP waveguide containing corrugated patterns at the edges of the wires. Except the edge pattern, the reference and the proposed device are identical in all respect, having line length of $L=150$ mm, wire width of $W=5$ mm,  inter-wire spacing of $S=5$ mm, and a ground plane underneath their substrates. For the engineered interconnect device, the corrugated groove length is $h=4$ mm, the groove width is $a=2$ mm, and the period of corrugation is $d=6$ mm. 

\section{Result And Discussion}

\subsection{Electromagnetic Insertion Loss}
We measured the attenuation constant (dB/cm) of a standalone meandered SSPP interconnect and a microstripline of the same footprint on a $2.6$ mm thick silicon substrate. Our goal was to analyze their electrical transmission properties and identify loss mechanisms. We investigated three scenarios: 1)  Si substrate with a perfect electric conductor, 2) Si substrate with copper, and 3) a $50$ $\mu$m silicon dioxide layer sandwiched between copper metal and the Si substrate (Fig.~\ref{fig:Geometry}c). Our analysis focused on frequencies up to $3$ GHz, typical for commercial processor chips, as our primary interest is base-band data transfer. At this frequency range, silicon's dielectric loss tangent is approximately $tan\delta=0.00187$ at 300 K, and copper's conductivity is $5.8\times10^7$ $S/m$\cite{b6}. 

\begin{figure}[ht]
    \centering
    \includegraphics[width=0.85\linewidth]{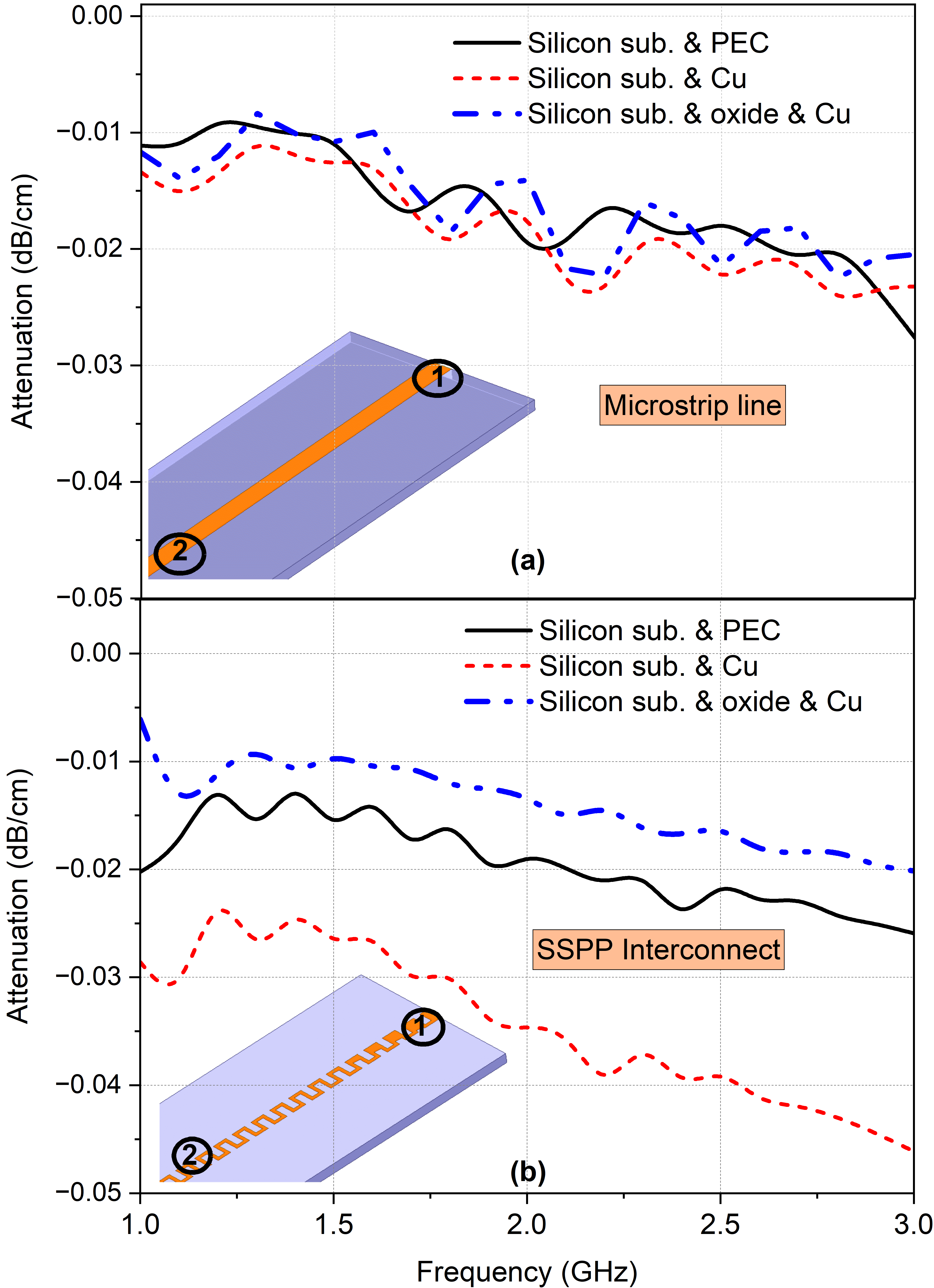}
    \caption{Attenuation characteristics of single channel interconnect for different substrate structures (a) Microstripline (b) meandered SSPP Interconnect}
    \label{fig:attenuation}
\end{figure}

The attenuation constant for the microstripline, shown in Fig.~\ref{fig:attenuation}(a), remains around $-0.015$ dB/cm, showing a slight degradation as frequency increases from $1$ to $3$ GHz. For all three tested cases (Si+perfect conductor, Si+Cu, and Si+SiO$_{2}$+Cu layer), the gross attenuation of microstripline is nearly identical. This indicates that losses in microstriplines primarily originate from the silicon substrate, with metallic losses in copper being negligible in this frequency range. The addition of a $50$ $\mu$m oxide layer doesn't improve signal loss because the silicon substrate is much thicker, and the microstrip mode's electric field largely remains within it, extending from the top metal to the ground. As shown in Fig.~\ref{fig:attenuation}(b), for the meandered SSPP waveguide with the same areal footprint, the attenuation constant for Silicon with perfect electric conductor case (Case 1) averages $-0.02$ dB/cm, which is a bit more degraded value than the microstripline. However, with Si+copper metal (case-2), the average attenuation increases to $-0.035$ dB/cm. This noticeable degradation in propagation in case-2 compared to case-1 suggests that copper loss becomes significant for SSPP wave propagation at $1$ GHz and above. This excess metallic loss is attributed to the slower group velocity of the SSPP signal, as evidenced by the slope of the SSPP waveguide's dispersion curve being two-third as less as that for the microstripline (Fig.~\ref{fig:dispersion}). To enhance propagation velocity and minimize metallic loss in SSPP waveguides, there are two main approaches: replacing the silicon substrate with a low-k dielectric or incorporating a thin low-k interfacial layer between the copper and silicon. The first option is thermally inefficient due to poor heat conduction of low-k materials, risking overheating. The second offers a better electrical-thermal trade-off. It retains the silicon substrate for better heat dissipation while lowering the dielectric constant near the metal surface, where the spoof plasmonic signal is highly concentrated. However, determining the optimal thickness for this oxide layer is crucial: too thin oxide cannot minimize copper loss, as the SSPP mode spills into the silicon; and too thick oxide would make heat dissipation worse. For our design, a $50$ $\mu$m thick oxide layer is found to significantly reduce attenuation, improving the average attenuation constant to $-0.015$ dB/cm within 1-3 GHz. As we'll show in Section~\ref{sec:thermal}, this specific oxide thickness minimally impacts the thermal advantages provided by the silicon substrate.

\subsection{Cross-talk Noise}\label{sec:cross-talk}
\begin{figure}[ht]
    \centering
    \includegraphics[width=0.85\linewidth]{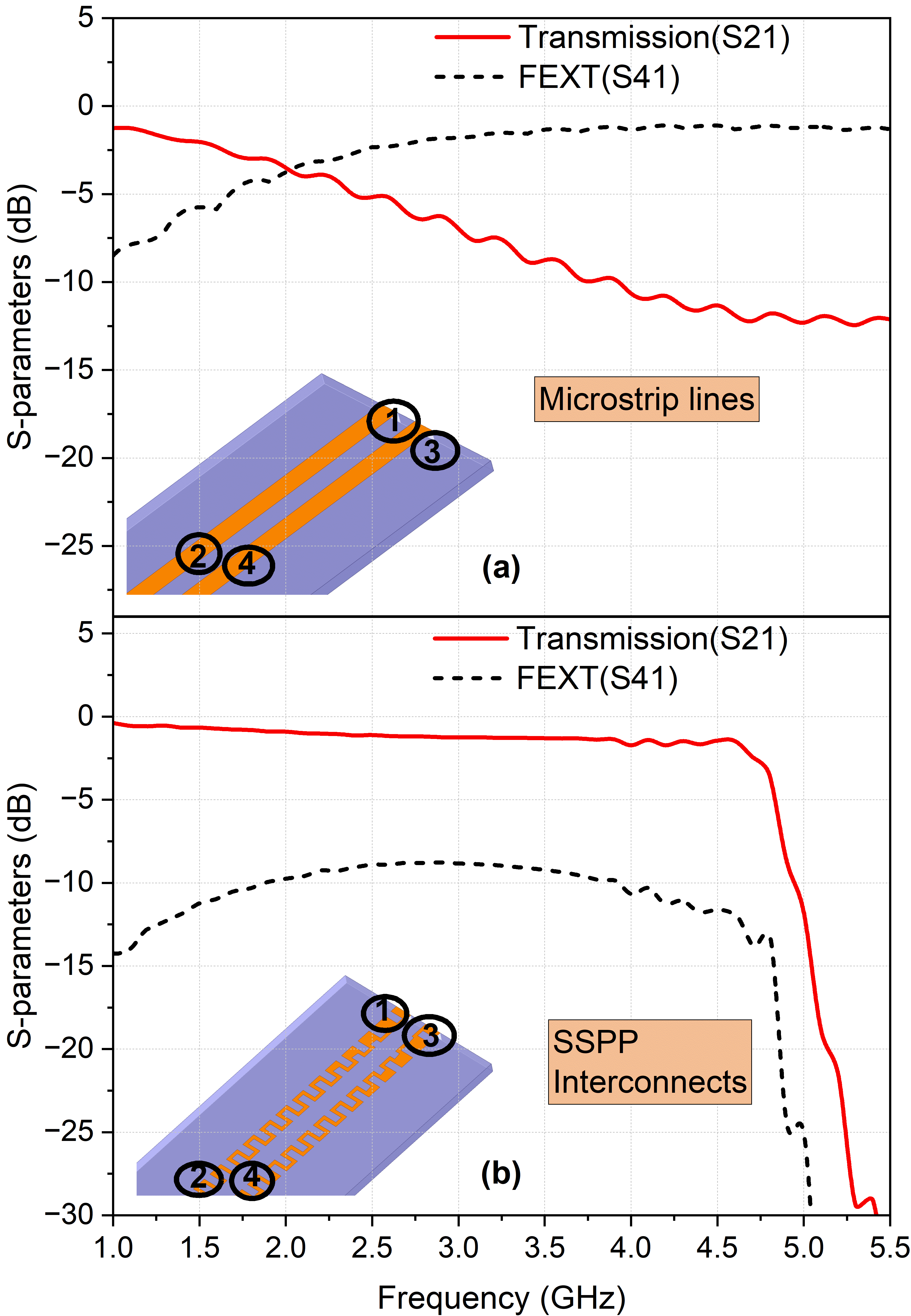}
    \caption{Transmission and Far-end Crosstalk Characteristics of the designed interconnect with Silicon Substrate and Copper wires with oxide layer in-between (a) Microstiplines (b) meandered SSPP Interconnect}
    \label{fig:S_param_Si_Cu_oxide}
\end{figure}

Crosstalk is a major bottleneck in high-density interconnects; our meandered SSPP waveguide mitigates this issue. In our 4-port device (Fig.~\ref{fig:Geometry}), ports 1 and 2 are for one interconnect, and ports 3 and 4 for another. We analyze direct transmission ($S_{21}$), near-end crosstalk ($S_{31}$), and far-end crosstalk (FEXT) ($S_{41}$). With $50$ $\Omega$ port termination and for good impedance matching, $S_{11}$ and $S_{31}$ remain low up to 5 GHz. However, close interconnect spacing and high substrate permittivity cause strong coupling, making FEXT dominate at high frequencies and reducing bandwidth below that of a standalone interconnect. Fig.~\ref{fig:S_param_Si_Cu_oxide}(a) shows that for adjacent microstriplines, crosstalk limits the bandwidth to 0--2 GHz; beyond this, $S_{41}$ exceeds $S_{21}$, causing the signal-to-crosstalk noise ratio to drop below 1. Conversely, Fig.~\ref{fig:S_param_Si_Cu_oxide}(b) demonstrates that replacing microstriplines with meandered SSPP interconnects significantly reduces FEXT ($S_{41}$) below $-10$ dB and boosts direct transmission ($S_{21}$) up to $5$ GHz. This $5$ GHz limit approximately corresponds to the spoof plasmonic resonance frequency. Thus, a pair of meandered SSPP waveguides offers a bandwidth of $0$--$5$ GHz, which is $\geq2.5\times$ as high as that of a pair of microstriplines. This bandwidth augmentation is achieved solely through geometric morph engineering of the metal, without requiring additional space, energy, or substrate material changes.

\begin{figure}[ht]
    \centering
    \includegraphics[width=0.85\linewidth]{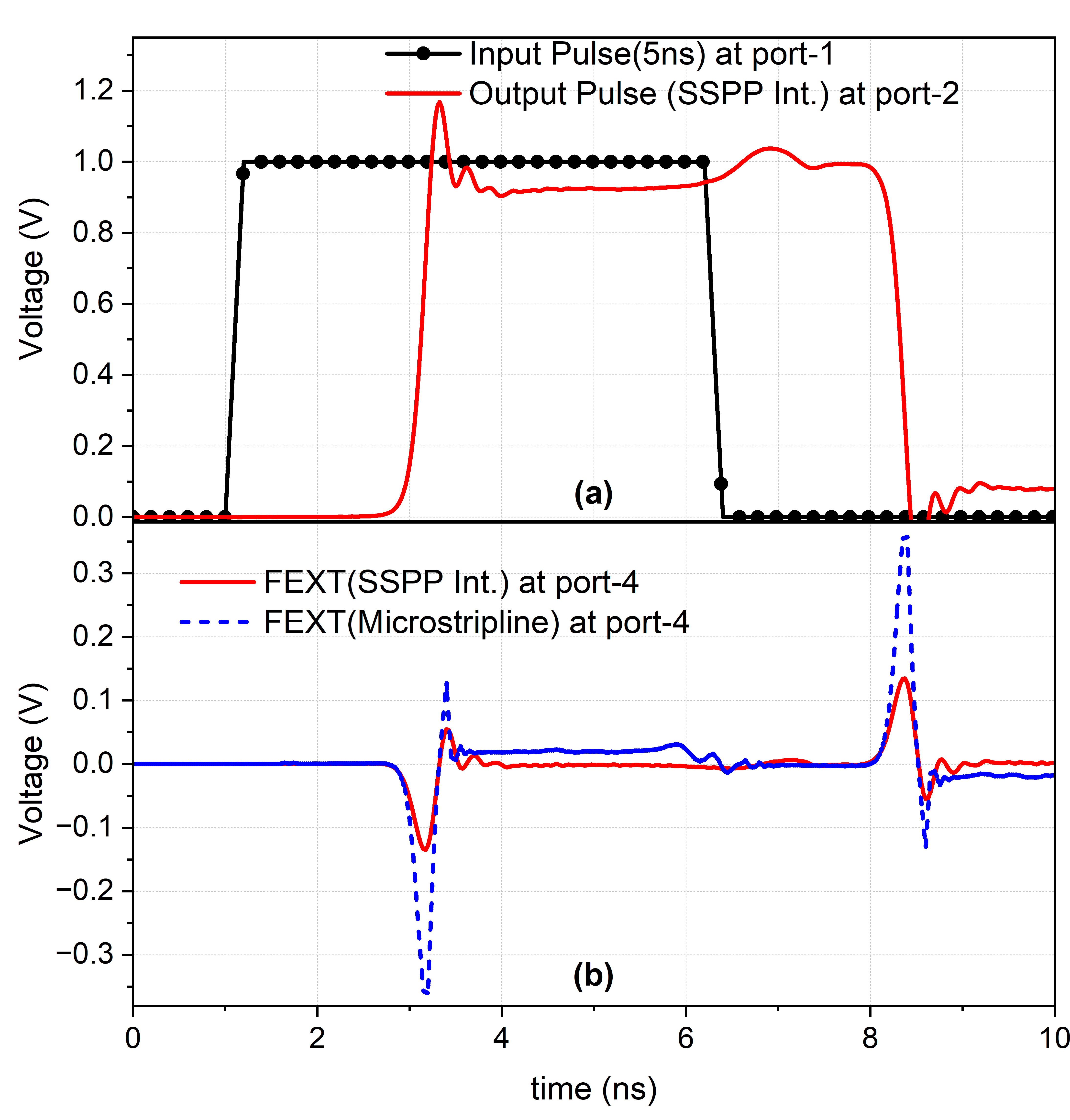}
    \caption{Time Domain Analysis for Microstrip Line and meandered SSPP (a) Comparison of Output Response at port-2 (b) Comparison FEXT behavior at port-4}
    \label{fig:time_domain_response}
\end{figure}

To assess the signal integrity of the proposed meandered SSPP waveguide, we injected a $5$ ns rectangular pulse into port-1 of an adjacent interconnect pair (Fig.~\ref{fig:Geometry}b). We then measured the direct transmission response at port-2 and the cross-talk at port-4. The output was reconstructed via convolution of impulse response of the system obtained from s-parameters. Fig.~\ref{fig:time_domain_response}(a) illustrates the output pulse at port-2 of the SSPP interconnect. It largely preserves the original rectangular shape with acceptable distortion, showing a propagation delay of $2$ ns for a $150$ mm long interconnect, thus $0.13$ $ns/cm$ delay. Fig.~\ref{fig:time_domain_response}(b) displays the output at port-4, which results from mutual coupling and exhibits noise spikes at the input pulse's edges. The meandered SSPP interconnect reduces these cross-talk spikes to one-third the level of microstrip lines (which reach $400$mV). This highlights its potential to enhance bandwidth and suppress cross-talk, thereby improving signal quality.

\subsection{Thermal Analysis}\label{sec:thermal} 
\begin{figure}[ht]
    \centering
    \includegraphics[width=0.85\linewidth]{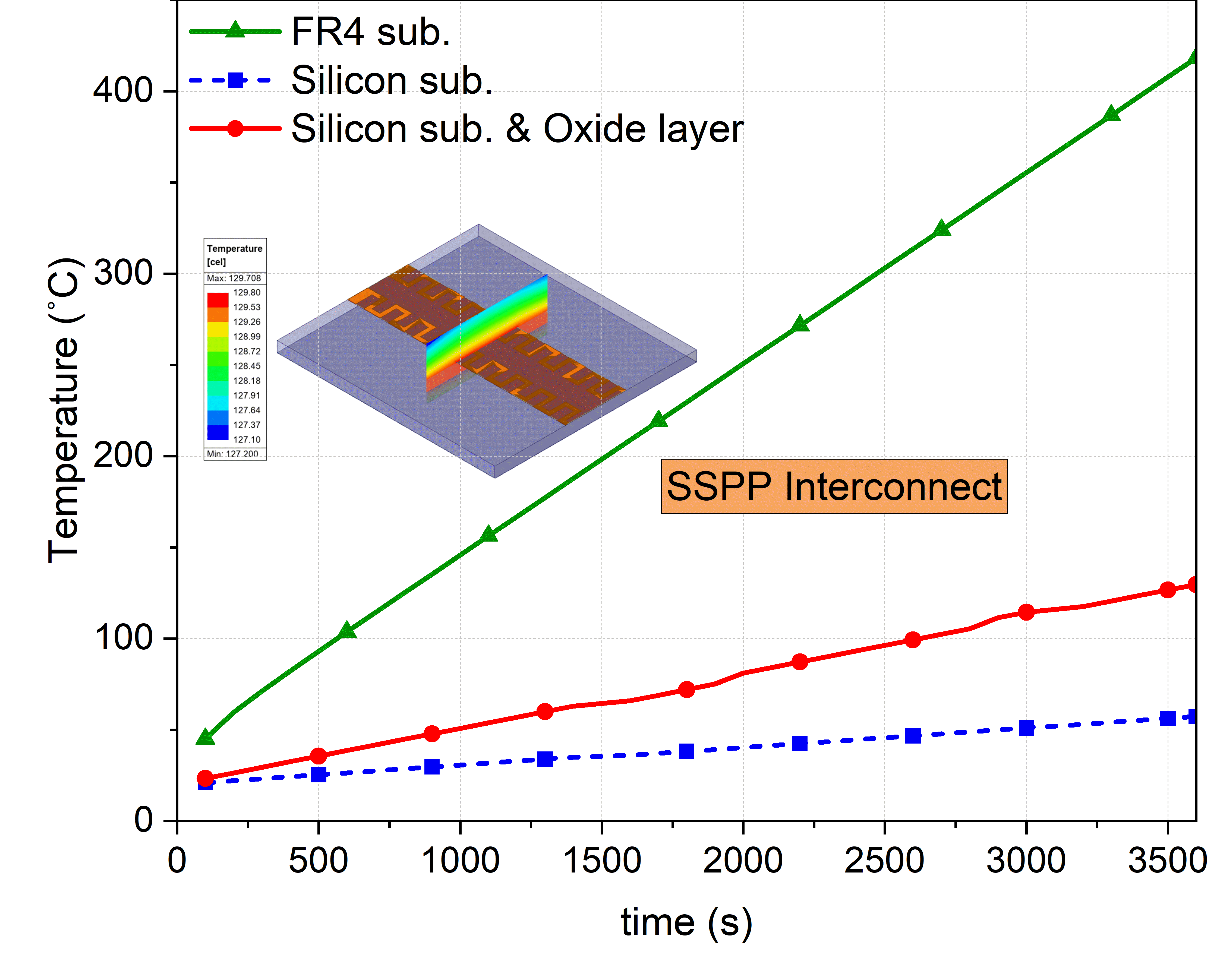}
    \caption{Thermal Analysis on SSPP wire with FR4 Substrate and Silicon Substrate with Oxide Layer. The inset shows the temperature distribution@1hour on Si substrate with oxide}
    \vspace*{-10pt}
    \label{fig:temperature_vs_time_SSPP}
\end{figure}

This section evaluates the impact of substrate materials on thermal performance. We simulated 1 Watt of continuous power absorbed by each SSPP interconnect on the top metal layer (Fig.~\ref{fig:Geometry}b), starting at an ambient temperature of $\SI{20}\celsius$ and observed the transient temperature for three different cases of choice of substrate: $2.6$ mm FR4, $2.6$ mm Silicon, and a $50$ $\mu$m silicon dioxide layer atop $2.6$ mm silicon. As Fig.~\ref{fig:temperature_vs_time_SSPP} shows, the top metal layer's temperature rises much slower with silicon, reaching $\SI{50}\celsius$ within an hour, whereas the FR4 substrate experience a temperature $8\times$ as high as that for silicon. This is due to silicon's significantly lower thermal resistance. Although adding a $50$ $\mu$m oxide layer resulted in a bit higher metal temperature than oxideless silicon ($\SI{110}\celsius$ after one hour), this temperature still remained considerably lower than that with FR4, demonstrating superior thermal management capabilities of the proposed substrate composition shown in Fig.~\ref{fig:Geometry}(c).

\section{Conclusion}
This work presents a compact, meandered spoof plasmonic (SSPP) interconnect on silicon substrate for high-bandwidth data transfer with enhanced electrical and thermal integrity. This structure reduces the footprint to $50\%$ of previously proposed mirror-symmetric double-sided SSPP waveguides. We addressed the issue of excess copper loss in SSPP interconnect by adding a $50$ $\mu$m $SiO_2$ layer between the copper and silicon layer. Crosstalk analysis reveals a reduction of FEXT noise floor by $\geq10$ dB and an increase of bandwidth by $\geq2.5\times$ compared to microstriplines, due to strong field confinement. Time-domain analysis shows a $5$ ns pulse propagates with a $0.13$ ns/cm delay in meandered SSPP, and crosstalk noise spikes are one-third that of microstriplines. Finally, the proposed meandered SSPP on oxide-coated silicon reduces temperature to one-fourth as low as that in low-k FR4 when dissipating $1$ W power, making it a promising candidate for next-generation chip-let interconnects.


\begin{thebibliography}{00}
\bibitem{b1} A. Sawada, Y. Kamiura, C. Fujikawa and O. Mikami,``Multi-channel Integration of Polymer Spot Size Expander on Silicon Photonics Chip'' 2024 IEEE CPMT Symposium Japan (ICSJ), Kyoto, Japan, 2024.
\bibitem{b2} S. Oh, Z. Zhang, G. Yan, P. K. Jo and M. S. Bakir, ``Heterogeneous Integration Enabled by 3-D Stitch-Chips'', in IEEE TCP-MT, vol. 15, pp. 113-122, Jan. 2025.
\bibitem{b3} F. Chen, J. Gill, D. Harmon, T. Sullivan and B. Li,``Measurements of effective thermal conductivity for advanced interconnect structures with various composite low-k dielectrics'',IEEE IRPS, USA, 2004, pp. 68-73.
\bibitem{b4} S. R. Joy, M. F. Bari, M. Z. Baten, F. Lan and P. Mazumder, ``A Reconfigurable Interconnect Technology based on Spoof Plasmon '', IEEE 19th International on Nanotechnology Macau SAR, Jul. 2019.
\bibitem{b5} R. Yasmin, M. J. Islam, I. Jahan, M. Z. Baten and S. Joy, ``Augmented Bandwidth of Interconnects by Hybridized Spoof Plasmonic Metamaterials '', IEEE 25th AP-S/URSI, Canada, Jul. 2025 in press.
\bibitem{b6} J. Krupka, J. Breeze, A. Centeno, and L. Jensen,``Measurements of Permittivity, Dielectric Loss Tangent and Resistivity of Float-Zone Silicon at Microwave Frequencies'', IEEE-TMTT, 2006, pp. 3995-4001.
\end{thebibliography}
\end{document}